\documentclass[]{aa}       
\usepackage{graphicx}
\usepackage{amssymb}
\usepackage{amsmath}
\usepackage{psfig}
\usepackage{longtable}
\usepackage{supertabular}

\begin{document}

\def\be{\begin{equation}}
\def\ee{\end{equation}}
\def\ba{\begin{eqnarray}}
\def\ea{\end{eqnarray}}
\def\d{\delta}
\def\e{\epsilon}
\def\f{\varphi}
\def\k{\varkappa}
\def\tde{\tilde}
\def\p{\partial}
\def\ms{\mathstrut}
\def\s{\strut}
\def\ds{\displaystyle}
\def\ts{\textstyle}
\def\b{\boldsymbol}
\def\r{\mathrm}
\def\G{\Gamma}
\def\sun{\odot}
\def\pks{PKS~2155$-$304}

\title{Large-scale flow dynamics and radiation in pulsar $\gamma$-ray binaries}

\author{
        V. Bosch-Ramon \inst{1} \and
        M.V. Barkov\inst{2,3}
    }

\institute{Dublin Institute for Advanced Studies, 31 Fitzwilliam Place, Dublin
2, Ireland; valenti@cp.dias.ie
\and
Max Planck Institut f\"ur Kernphysik, Saupfercheckweg 1, Heidelberg 69117,
Germany
\and
Space Research Institute, 84/32 Profsoyuznaya Street, Moscow, Russia
}

\titlerunning{Large-scale flow dynamics and radiation in pulsar $\gamma$-ray binaries}

\offprints{V. Bosch-Ramon; \email{valenti@cp.dias.ie}}

\date{Received <date> / Accepted <date>}

\abstract
{Several gamma-ray binaries show extended X-ray emission that may be associated
to interactions of an outflow with the medium. 
Some of these systems are, or may be, high-mass binaries harboring young
nonaccreting pulsars, in which
the stellar and the pulsar winds collide, generating a powerful outflow that
should terminate at some point 
in the ambient medium.} 
{This work studies the evolution and termination, as well as the related radiation, 
of the shocked-wind flow generated in high-mass binaries hosting powerful pulsars.} 
{A characterization, based on previous numerical work, is given for the
stellar/pulsar wind interaction. Then, an analytical study of the
further evolution of the shocked flow and its dynamical impact on the
surrounding medium is carried out. 
Finally, the expected nonthermal emission from the flow termination shock,
likely the dominant emitting region, is calculated.}
{The shocked wind structure, initially strongly asymmetric, becomes a
quasi-spherical, 
supersonically expanding bubble, with its energy coming from
the pulsar and mass from the stellar wind. 
This bubble eventually interacts with the environment on $\sim$~pc scales, 
producing a reverse and, sometimes, a forward shock. Nonthermal leptonic radiation 
can be efficient in the reverse shock. Radio emission is expected to be faint, whereas
X-rays can easily reach detectable fluxes. Under very low magnetic fields and large 
nonthermal luminosities, gamma rays may also be significant.}
{The complexity of the stellar/pulsar wind interaction is likely to be smoothed out outside the binary
system, where the wind-mixed flow accelerates and eventually terminates in a
strong reverse shock. This shock may be behind the extended X-rays observed in some binary
systems. For very powerful pulsars, part of the unshocked pulsar wind may directly 
interact with the large-scale environment.}

\maketitle

\section{Introduction}

Gamma-ray binaries are compact sources that present nonthermal emission in the GeV and/or TeV bands and, typically, they are also moderate-to-strong emitters in radio and X-rays. The
radiation in gamma-ray binaries is related to transient or persistent outflows, which can interact either with themselves in the form of internal shocks or with the environment on different
scales. The number of members of the class of gamma-ray binaries is growing, being at present around ten, including unconfirmed candidates (e.g. Albert et al. \cite{alb06,alb07}; Aharonian
et al. \cite{aha05a,aha05b}; Abdo et al. \cite{abd09a,abd09b,abd09c,abd10a,abd10b,abd11};Tavani et al. \cite{tav09a,tav09b}; Sabatini et al. \cite{sab10}; Williams et al. \cite{wil10}; Tam
et al. \cite{tam10}; Hill et al. \cite{hil11}; Falcone et al. \cite{bon11}; Corbet et al. \cite{cor11}). 

Several gamma-ray binaries present hints or clear evidence of interactions between an outflow and their medium on large
scales: Cygnus~X-3, LS~I~+61~303, LS~5039, PSR~B1259$-$63, and the gamma-ray emitter candidate Cygnus X-1 (Heindl et al.
\cite{hei03}; S\'anchez-Sutil et al. \cite{san08}; Paredes et al. \cite{par07}; Pavlov et al. \cite{pav11a}; Durant et al.
\cite{dur11}; Mart\'i et al. \cite{mar96}; Gallo et al. \cite{gal05}; Russell et al. \cite{rus07}). 

Cygnus~X-1 and Cygnus~X-3 are high-mass microquasars, in which accretion leads to jets that can interact with the ISM (e.g.
Vel\'azquez \& Raga \cite{vel00}; Heinz \cite{hei102}; Heinz \& Sunyaev \cite{hei202}; Bosch-Ramon et al. \cite{bos05};
Zavala et al. \cite{zav08}; Bordas et al. \cite{bor09}; Bosch-Ramon et al. \cite{bos11}). On the other hand, PSR~B1259$-$63
is a system formed by an O9.5\,V star and a nonaccreting millisecond pulsar (Johnston et al. \cite{joh92}; Negueruela
et al. \cite{neg11}). Instead of coming from a jet, like in microquasars, the nonthermal emission of this source is thought
to originate in the region where the star and the pulsar winds collide (Tavani \& Arons \cite{tav97}), with its radio
emission extending far away from the binary (Mold\'on et al. \cite{mol11a}). 
PSR~B1259$-$63 also presents extended X-ray emission on scales of $\sim 4-15"$ (i.e. a projected linear size of $\sim 1.5-5\times
10^{17}$~cm at the source distance of 2.3~kpc; Negueruela et al. \cite{neg11}), with a flux $\sim
10^{-13}$~erg~s$^{-1}$~cm$^{-2}$ and photon index $\sim 1.6$ (Pavlov et al. \cite{pav11a}). The extended radio and X-ray
emission would be produced in different regions but still in the shocked stellar-pulsar wind structure, which propagates away
from the binary and should eventually terminate in the external medium. 

The nature of LS~5039 and LS~I~+61~303 is still unknown due to the lack of clear pulsar or accretion features, conclusive
system dynamical information, or nonthermal emission evidence (e.g. Casares et al. \cite{cas05a}, Casares et al.
\cite{cas05b}, Sarty et al. \cite{sar11}; Rea et al. \cite{rea10,rea11}, McSwain et al. \cite{mcs11}, Rib\'o \cite{rib11}; 
see also 
Bosch-Ramon \& Khangulyan \cite{bos09}). In the past, several works have considered LS~5039 and LS~I~+61~303 good microquasar candidates because of their
radio morphology and emission properties (e.g. Taylor et al. \cite{tay92}; Paredes et al. \cite{par00}; Massi et al.
\cite{mas01,mas04}; Paredes et al. \cite{par02}; Romero et al. \cite{rom05}; Paredes et al.
\cite{par06}; Bosch-Ramon et al. \cite{bos06}; Massi \& Kaufman Bernad\'o \cite{mas09}). However, it has also been proposed
that the phenomenology of these sources and, in particular, their milliarcsecond (mas) scale radio morphology may be more
compatible with a nonaccreting pulsar scenario (e.g. Maraschi \& Treves \cite{mar81}, Martocchia et al. \cite{mar05}, Dubus
\cite{dub06}, Chernyakova et al. \cite{che06}, Sierpowska-Bartosik \& Torres \cite{sie08}, Sierpowska-Bartosik \& Torres
\cite{sie09}; see however Romero et al. \cite{rom07}). Both binaries are high-mass systems, with LS~5039 harboring an O6.5\,V
star (Clark et al. \cite{cla01}), and LS~I~+61~303, an early Be star (Hutchings \& Crampton \cite{hut81}). 

Extended X-rays have been found in LS~5039, with angular size $\theta\sim 1'-2'$ ($\sim 2-4\times 10^{18}$~cm at
2.5~kpc; Casares et al. \cite{cas05a}), X-ray flux $\approx 9\times 10^{-14}$~erg~s$^{-1}$~cm$^{-2}$ and photon index $\sim
1.9-3.1$ (Durant et al. \cite{dur11}), and possibly also in LS~I~+61~303, with $\theta\sim 10"$ ($\sim 3\times 10^{17}$~cm at
2~kpc; Frail \& Hjellming \cite{fra91}), and X-ray flux $\approx 2\times 10^{-14}$~erg~s$^{-1}$~cm$^{-2}$ (Paredes et al.
\cite{par07}\footnote{Rea et al. \cite{rea10} did not find evidence of extended X-rays in LS~I~+61~303 in longer observations
than those studied in Paredes et al. \cite{par07}. However, in their observations the X-ray counts were integrated along one
spatial dimension in the CCDs, which only allows finding an extension signal in specific directions.}). It is noteworthy
that the gamma-ray binaries HESS~J0632$+$057 (Aharonian et al. \cite{aha07}; Hinton et al. \cite{hin09}; Skilton et al.
\cite{ski09}; Falcone et al. \cite{fal10,bon11}; Mold\'on et al. \cite{mol11b}) and 1FGL~J1018.6$-$5856 (Corbet et al.
\cite{cor11}; Pavlov et al. \cite{pav11b}; de Ona Wilhelmi et al. \cite{deo10}) may host a nonaccreting
pulsar as well, although a microquasar scenario cannot be discarded. 

Summarizing, among the several gamma-ray binaries presenting hints or evidence of interactions with the medium, three of them
may (one for sure) host a nonaccreting pulsar. In addition, extended X-ray emission may be eventually discovered as well in 
HESS~J0632$+$057 and 1FGL~J1018.6$-$5856, another two pulsar binary 
candidates.
At present, the dynamics and radiation of an outflow from a pulsar gamma-ray
binary has not been studied far from the binary system. Given the recent findings of extended emission, it seems necessary to
analyze the flow evolution beyond the stellar/pulsar wind region and its interaction with the ISM. This
is the goal of this work, which is distributed as follows. In Sect.~\ref{phys}, the main properties of the binary generated
flow are described. In Sect.~\ref{dyn}, the general properties of the interaction between the flow and the external medium,
the progenitor supernova remnant (SNR; young sources), or the interstellar medium (ISM; older sources) are characterized
analytically, and in Sect.~\ref{nt}, the expected nonthermal emission from radio to gamma rays is computed. Finally, in
Sect.~\ref{disc}, the results are discussed in the context of PSR~B1259$-$63, LS~I~+61~303 and LS~5039 (assuming
that the last two host a powerful pulsar), and some final remarks are made in Sect.~\ref{rem}. All through the paper, cgs
units are used.

\section{Physical scenario}\label{phys}

\subsection{The stellar-pulsar wind shock}

\begin{figure*}[]
   \centering
\includegraphics[angle=0, width=0.9\textwidth]{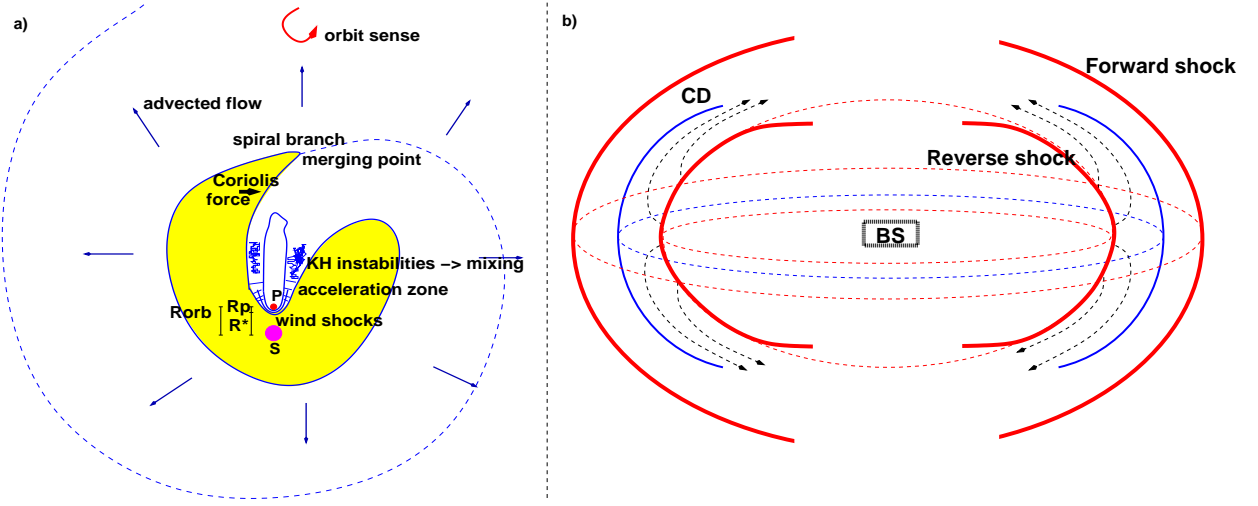}
\caption{{\bf a)} Sketch of the pulsar high-mass binary scenario on scales larger than the binary
system. The star (S) and pulsar (P) shocked winds 
trace a spiral-shape due to the Coriolis force and the pulsar orbital motion.
Eventually, both shocked winds mix due to Kelvin-Helmholtz instabilities, and
the
spiral arms merge. 
{\bf b)} Cartoon of the 
bubble (case $\eta<1$) driven by the shocked flows formed within the binary system (BS) and 
made of pulsar- and stellar-wind material. The bubble
expands and accelerates, eventually terminating in a shock where its 
ram pressure is equal to the pressure of the SNR or the shocked ISM for old
enough sources. This occurs at the contact discontinuity (CD).}
\label{f0}
\end{figure*}

In Figure~\ref{f0}, the pulsar high-mass binary scenario is sketched. For a general description of the scenario on the binary
scales, see for instance Tavani \& Arons (\cite{tav97}).
As seen in the figure, the stellar and the pulsar winds collide to form two bow-shaped
standing shocks. An important parameter that characterizes the wind
interaction region is the pulsar to stellar wind momentum flux ratio (e.g. Bogovalov et al. \cite{bog08}): 
\be
\eta=\frac{L_{\rm sd}}{\dot{M}_{\rm w}v_{\rm w}c}\approx 0.06\,L_{\rm sd37}\dot{M}^{-1}_{\rm w-6.5}v^{-1}_{\rm w8.5},
\label{eta}
\ee
where $\dot{M}_{\rm
w-6.5}=(\dot{M}_{\rm w}/3\times 10^{-7}\,M_\odot\,{\rm yr}^{-1})$ is the stellar
mass-loss rate, $v_{\rm w8.5}=(v_{\rm
w}/3\times 10^8\,{\rm cm~s}^{-1})$ is the stellar wind velocity, and $L_{\rm
sd37}=(L_{\rm sd}/10^{37}\,{\rm erg~s}^{-1})$ is the pulsar spin-down
luminosity, taken here as equal to the
kinetic luminosity of the pulsar wind. The stellar wind is assumed isotropic.
The pulsar orbital velocity, of a few hundred km~s$^{-1}$, 
is much lower than $v_{\rm w}$, typically $\sim 2\times 10^8$~cm~s$^{-1}$ for massive stars (e.g. Puls et al. \cite{pul09}), 
so it has been neglected in this section. However, the orbital velocity may play a relevant role in 
determining $\eta$ in some cases, e.g. in the presence of a stellar decretion disk, 
as discussed in Sect.~\ref{eqfl}. It is also noteworthy that the
pulsar wind is simplified here as isotropic, although a more refined treatment should account for anisotropy 
(see, e.g., Bogovalov \& Khangoulian
\cite{bog02}, Bogovalov et al. \cite{bog11}).
 
For reasonable $\eta$-values $<1$ (a scenario with $\eta>1$ is briefly discussed in Sect.~\ref{pulsd}), the
shocked wind structure is dominated by the stellar wind ram pressure, and points
away from the star on the
binary scales. 
The size of the wind colliding region can be
characterized by the distance between the two-shocks contact discontinuity (CD)
and the pulsar, 
\be
R_{\rm p}=\frac{\sqrt{\eta}\,R_{\rm orb}}{(1+\sqrt{\eta})}\,,
\label{rp}
\ee
which is $\approx 7\times10^{11}$~cm for $\eta=0.1$ and 
a distance between the star and the pulsar of $R_{\rm orb}=3\times 10^{12}$~cm. 
The distance between the CD and the star 
is then $\approx 2.3\times 10^{12}$~cm. The
CD, which separates the shocked stellar and pulsar winds, has an asymptotic
half-opening angle (see Bogovalov et al. \cite{bog08} and references therein):
\be
\alpha\approx 28.6^\circ\,(4-\eta^{2/5})\,\eta^{1/3}\,.
\ee 
Whereas the main contribution of momentum
an mass generally comes from the stellar wind, the energy is mostly provided by
the pulsar wind, as seen from
the pulsar to stellar wind luminosity ratio: 
\be
\frac{L_{\rm sd}}{L_{\rm w}}=\frac{2\,L_{\rm sd}}{\dot{M}_{\rm w}v_{\rm w}^2}\approx 12\,L_{\rm sd37}\dot{M}^{-1}_{\rm w-6.5}v_{\rm w8.3}^{-2}\,. 
\ee

We notice that the wind of the pulsar prevents the latter to accrete from the stellar wind for a wide range of $\eta$-values. As a conservative estimate, we 
can derive the $\eta$-value for which $R_{\rm p}$ is equal to the gravitational capture radius (Bondi \& Hoyle \cite{bon44}):
\be
\eta\approx \left(\frac{2GM}{(R_{\rm orb}v_{\rm w}^2)^2}\right)\sim 2\times10^{-6} R_{\rm orb12.5}^{-2}\,v_{\rm w8.5}^{-4}\,,
\ee
where $M$ is the pulsar mass, and $R_{\rm orb12.5}=(R_{\rm orb}/3\times 10^{12}\,{\rm cm})$. 

\subsection{Shocked flow evolution}\label{sfe}

After being shocked, the pulsar wind facing the star moves at a speed $c/3$, but
suffers a strong reacceleration along the CD produced by a strong pressure
gradient, related to the
anisotropy and inhomogeneity of the stellar wind at the pulsar location
(Bogovalov et al. \cite{bog08}). Because of the very low density in the downstream
region, the pulsar wind shock is
adiabatic (Bogovalov et al. \cite{bog08}; Khangulyan et al. \cite{kha08}). The
stellar wind shock could be either adiabatic or radiative, but in either case
the stellar wind velocity will be $\ll c$. Therefore, the two shocked winds 
have a relative velocity $\sim c/3-c$, depending on
how much the shocked pulsar
wind has reaccelerated. Such a large velocity difference is likely to lead to
the development of Kelvin-Helmholtz (KH) instabilities in the CD. This will
produce entrainement of shocked
stellar wind into the lighter and faster shocked pulsar wind.   If the stellar
wind shock is radiative, then the matter downstream collapses into a dense and
cool thin layer, diminishing
even farther the stability of the whole shocked structure (e.g. Stevens et al.
\cite{ste92}; Pittard \cite{pita09}). Perturbations generated in the interface
will initially move with the
stellar shocked wind, growing on a timescale $t_{\rm KH}\sim l/c_{\rm sw}$, where $l$ is
the size of the perturbation, and $c_{\rm sw}$ the sound speed of the shocked
stellar wind. Taking $c_{\rm
sw}\sim v_{\rm w}\sim 10^8$~cm~s$^{-1}$ and $l\sim R_{\rm p}=10^{12}$~cm, one obtains $t_{\rm
KH}\sim 10^4$~s, approximately the lapse needed by the perturbations to enter in
the nonlinear regime. This
implies that instabilities, turbulence, and mixing will develop on spatial scales that are similar
though slightly larger than those of the binary system. 

We now discuss the impact of the orbital motion on the shocked winds. To do this, we focus on three cases:
the possibly more common 
$\eta<1$, (e.g. $\dot{M}_{\rm w}>2.5\times 10^{-8}\,M_\odot$~yr$^{-1}$ for $v_{\rm w}\sim 2\times 10^8$~cm~s$^{-1}$ and $L_{\rm sd}\sim
10^{37}$~erg~s$^{-1}$); more briefly, the extreme case $\eta>1$; and under the presence of a stellar decretion disk.

\subsubsection{Stellar wind momentum flux dominance}\label{stardom}

We analyze the dynamics of the gas flow in the coordinate system that corotates 
with the binary system at an angular velocity $\Omega=2\pi/T$, where $T$ is the orbital 
period. The X-axis coincides 
with the line connecting the centers of the stars, the Y-axis lies in the orbital plane, and
the Z-axis is normal to the orbital plane. In such a coordinate system, the Coriolis 
acceleration is $\b a_c=-2\,\b \Omega\times \b v$. The stellar wind
moves along the X-axis with speed $\sim v_{\rm w}$, acquiring a
velocity $v_{\perp}=2\,\Omega\,v_{w}\,t$ in the Y-direction. In the frame of the 
shocked-wind flow,
$t=x/v_{\rm w}$, and $x$ is the distance, along the X-axis, 
between the pulsar and a point farther away from the star.
The dynamical pressure of the 
stellar wind in the Y-direction can be estimated as 
$P_{\perp}\approx \rho_{\rm w} v_{\perp}^2\approx (2\Omega t)^2\rho_{\rm w} v_{\rm w}^2
\approx (2\Omega t)^2 P_{\rm w}$ ($P_{\rm w}=\rho_{\rm w} v_{\rm w}^2$). 
From this, the location where the 
pulsar wind total pressure is balanced by the stellar wind can be
derived with
\be
\frac{L_{\rm sd}}{4\pi c x^2}=P_{\perp}=\rho_{\rm w}\,\left(\frac{4\pi}{T}\right)^2\,x^2\,,
\label{perbal}
\ee 
where $\rho_{\rm w} = \rho_{\rm w0}/(1+x/R_{\rm orb})^2$, with
$\rho_{\rm w0} = \dot{M}_{\rm w}/4\pi R_{\rm orb}^2 v_{\rm w}$ 
the stellar wind density at the pulsar location. For $x\ll R_{\rm orb}$,
Eq.~(\ref{perbal}) can be simplified to 
\be
x=\left(
\frac{L_{\rm sd}v_{\rm w}T^2R_{\rm orb}^2}{(4\pi)^2c\dot{M}_{\rm w}}\right)^{1/4}\,;
\label{lsmall}
\ee
and for $x\gg R_{\rm orb}$,
\be
x=\left(
\frac{L_{\rm sd}v_{\rm w}T^2}{(4\pi)^2c\dot{M}_{\rm w}}\right)^{1/2}\,.
\label{lbig}
\ee
For typical parameters, the solution is closer to the one given by Eq.~(\ref{lbig}), therefore the 
bending distance can be written as
\be
x_0\approx 7\times 10^{12}\,L_{\rm sd37}^{1/2}\,v_{\rm w8.5}^{1/2}\,T_{6}\,\dot{M}^{-1/2}_{\rm w-6.5}\,{\rm cm}\,,
\ee 
where $T_{6}=(T/10^6\,{\rm s})$.
This distance is much less than in the purely ballistic case, for which
$x_0\sim c\,T=3\times 10^{16}\,T_{6}$~cm.

It is noteworthy that the pulsar wind is shocked even in the direction opposite to the star, 
within a distance $\sim x_0$. Therefore, a significant fraction of the pulsar wind luminosity
is reprocessed outside the binary system. Interestingly, this region may dominate the very high-energy output in
pulsar gamma-ray binaries with high photon-photon absorption (see the discussion in Bosch-Ramon et al. \cite{bos08}).

If KH instabilities by themselves have not already mixed the shocked winds,
isotropizing their particle, momentum, and energy fluxes, together with the Coriolis force 
they very likely will. Eventually, the shocked-wind flow
will probably end up as a trans- or subsonic turbulent bubble, loaded with stellar wind
mass and pulsar wind energy. The bubble is not confined because there is still a strong pressure gradient
radially outwards. Therefore, and assuming that the whole process is adiabatic, 
any thermal, turbulent, or magnetic energy will eventually
end up in the form of radial bulk motion. The role of the magnetic field should
not change this
picture much, unless the initial stellar and pulsar winds had a dynamically dominant
magnetic field. (For the role of the magnetic field on binary scales; 
see Bogovalov et al. \cite{bog11}.) The maximum velocity expansion of the bubble will be roughly its
initial sound speed, which
for a maximum entropy initial state (i.e. right after isotropization) is
\be
v_{\rm exp}\sim\sqrt{\frac{2L_{\rm sd}}{\dot{M}_{\rm w}}}\approx 10^9\,L^{1/2}_{\rm sd37}\,\dot{M}^{-1/2}_{\rm w-6.5} \mbox{ cm s}^{-1}.
\label{vexp}
\ee 
Typically, the mas radio emission in pulsar massive binaries will come from scales larger than $x_0$, and thus 
this radiation should be strongly affected by the Coriolis force, as well as by shocked-wind mixing.
The situation described
in this section is pictured in Fig.~\ref{f0} (left).

\subsubsection{Pulsar wind momentum flux dominance}\label{pulsd}

\begin{figure}[]
   \centering
\includegraphics[angle=0, width=0.53\textwidth]{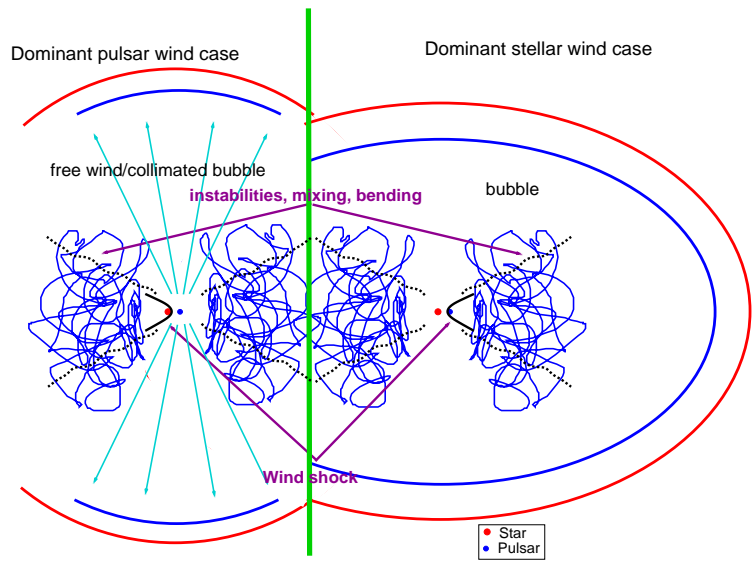}
\caption{Sketch of the two possibilities for the outcome of the interaction between the pulsar and 
the star winds, in the context of a binary system. To the right, the case for a very powerful pulsar, with $\eta>1$, 
is presented. To the left, the case with $\eta<1$.}
\label{fig2cas}
\end{figure}

For a dominant pulsar wind, the shocked structure closes towards the star. In this case, the Coriolis force exerted by
the light pulsar wind is too low and the shocked structure moves ballistically, with a bending typical distance $\sim v_{\rm
w}\,T$. Along the orbit, however, the natural bending of the shocked structure directly exposes it to the pulsar wind ram pressure, which pushes
and opens the spiral to some degree. As for $\eta<1$, for $\eta>1$ the instabilities in the CD will also take place, and the
formation of a mixing layer is expected to occupy part of the region of the free pulsar wind. However, in the direction
perpendicular to the orbital plane, the pulsar wind will likely propagate freely. The energetics of this free wind may be
significant, eventually terminating as a jet-like structure interacting with the surrounding medium (see, e.g., Bordas et al.
\cite{bor09}). A sketch of the two cases discussed, with $\eta>1$ and $<1$, is shown in Fig.~\ref{fig2cas}.

\subsubsection{The equatorial flow case}\label{eqfl}

The situation discussed so far is that of two spherical winds interacting with each other. However, some of the systems
considered in this work host Be stars, which present decretion disks around the equator.
These disks are thought to be much denser than the radiatively driven stellar wind. The disk flow is quasi-Keplerian, and
moves subsonically in the radial direction with velocities around $\sim 1$~km~s$^{-1}$ (e.g. Porter \& Rivinius
\cite{por03}). Given the slow flow velocity, the flow speed determining the flux momentum ratio, in the pulsar reference frame, is near the orbital
velocity of the pulsar. Outside the equatorial disk, the radiatively driven wind in Be stars is
most likely polar. Such a configuration of the two stellar flows probably makes the geometry of the colliding wind region
quite complex, enhancing the development of instabilities in the shocked-flow contact discontinuity (for 
simulations of this scenario on binary system scales, see Romero et al. \cite{rom07} and Okazaki et al. \cite{oka11}). The shocked disk material has a
much larger density than the light polar wind. This implies that development of instabilities may be slower, although the
sound speed of the shocked disk will be similar to the pulsar's orbital velocity, the dynamical speed in this case. The high
density of the shocked equatorial flow implies that it will likely be radiative, thereby potentiating instabilities and
fragmentation. Finally, the mass-loss rate of the disk is similar or even smaller than that of the polar wind, and thus once
accelerated by the pulsar wind, the flow evolution on the largest scales should not be very different from the isotropic wind
case. The dense and fragmented shocked disk material will likely introduce inhomogeneities into the larger scale shocked flow,
although eventually the isotropization of the particle, momentum, and energy fluxes of the larger scale structure seems
a reasonable guess. However, hydrodynamical simulations on larger scales than the binary are mandatory for validating
the assumption.

\subsection{Bubble large-scale evolution}

The pulsar started its life after a supernova explosion.
For an explosion
energy $E_{\rm SNR}\sim
10^{51}$~erg and ISM density $\rho_{\rm ism}=1.7\times 10^{-23}$~g~cm$^{-3}$
($n_{\rm ism}=10$~cm$^{-3}$), the SNR becomes radiative (Blinnikov et al. \cite{biu82}) 
after $t_{\rm c}\approx 2\times 10^4$~yr when achieving the 
radius $R_{\rm c}\approx 11\,(E_{\rm SNR51}/n_{\rm ism1})^{1/3}$~pc, 
where $E_{\rm SNR51}=(E_{\rm SNR}/10^{51}\,{\rm erg})$ and $n_{\rm ism1}=(n_{\rm ism}/10\,{\rm cm}^{-3})$. 
In this context, the binary system will leave the SNR after a time 
\be
t_{\rm sp}\approx \left(\frac{2.8\,E_{\rm SNR}\,R_{\rm c}^2}{\rho_{\rm ism}}\right)^{1/5}\,v_{\rm p}^{-7/5}\,;
\label{tsps}
\ee
i.e., $t_{\rm sp}\approx 8\times 10^4$~yr for 
$v_{\rm p}\approx 2\times 10^7$~cm~s$^{-1}$. 

Inside the SNR, given the high sound speed of the
ambient medium, the binary driven bubble cannot generate a forward shock. Once
in the ISM, the bubble forms a slow forward shock. In
both situations, the supersonic expanding bubble suffers a strong reverse shock
to balance the external pressure,
either from the hot SNR ejecta or the shocked ISM. This reverse shock is expected to
have a speed $\sim v_{\rm exp}$ in the bubble reference frame. If the pulsar 
is powerful enough, the reverse
shock can power a moderately bright nonthermal emitter, with a potential 
nonthermal luminosity $L_{\rm nt}$ as high as $L_{\rm sd}$. This estimate 
is done by neglecting the radiation losses on the binary system scales, but
as long as adiabatic losses are likely to dominate in the colliding wind region
(e.g. Khangulyan et al.
\cite{kha08}), this approximation should suffice at this stage. Thermal emission from
the reverse shock can be discarded, given the low densities. Some extended thermal
emission will be generated in the
shocked ISM, but with its peak in the ultraviolet it may be hard to detect.
Figure~\ref{f0} (right) sketches the bubble termination region.  

\section{Dynamics of the bubble interacting with the medium}\label{dyn}

The interaction between the bubble and its environment takes place in three steps.
At early times, the bubble is still surrounded by the SNR in its adiabatic 
or Sedov phase (Sedov \cite{sed59}). Later on, the SNR forward shock in the ISM 
becomes radiative, entering into the so-called snow-plow phase (Cox \cite{cox72}; Blinnikov et al. \cite{biu82}). 
Finally, after the binary leaves the SNR or the latter dissipates, the bubble interacts 
directly with the ISM.

In the adiabatic regime, the SNR properties can be characterized through 
the SNR/ISM forward shock radius (Sedov \cite{sed59})
\be
R_{\rm SNR}\approx 1.15\,(E_{\rm SNR}/\rho_{\rm ism})^{1/5}\,t_{\rm
p}^{2/5}\,,
\ee
where $t_{\rm p}$ is the pulsar age, velocity
\be
v_{\rm SNR}\approx (2/5)\,R_{\rm SNR}/t_{\rm p}\,, 
\ee
and the inner pressure is
\be 
P_{\rm SNR}\approx 
\frac{E_{\rm SNR}}{2\pi R_{\rm SNR}^3}\,, 
\label{psnr}
\ee
assumed here to be constant.
Equilibrium between
bubble and SNR pressures
determines the location of the boundary between these two regions: 
$R_{\rm bcd}\approx (3\,L_{\rm sd}\,t_{\rm p}/10\pi\rho_{\rm
ism}\,v_{\rm b}^2)^{1/3}$,
and the equilibrium
between bubble preshock and postshock total pressures determines the
location of the bubble reverse
shock: 
\be
R_{\rm brs}\approx \sqrt{\frac{L_{\rm sd}}{2\pi P_{\rm SNR}v_{\rm exp}}}\,.
\label{brss}
\ee 
Since the hot ejecta region has a high sound velocity,
the bubble forward shock inside the
SNR quickly becomes a sound wave, and the bubble shape
becomes spherical. 

For $t>t_{\rm c}$, the SNR enters into the radiative phase, so 
the snow-plow solution has to be used (Blinnikov et al. \cite{biu82}). The SNR/ISM 
forward shock has now a radius of
\be
R_{\rm SNR}\approx\left(\frac{2.8\,E_{\rm SNR} R_{\rm c}^2 t^2}{\rho_{\rm ism}} \right)^{1/7}\,,
\label{Rsps}
\ee
a velocity of
\be
v_{\rm SNR}=(2/7)\,R_{\rm SNR}/t_{\rm p}\,,
\ee
and an inner pressure of
\be
P_{\rm SNR}\approx\frac{1}{2\pi}\frac{\epsilon E_{\rm SNR} R_{\rm c}^2}{R_{\rm SNR}^5}\,,
\label{psps}
\ee
where $\epsilon=0.24$.
As before, the bubble reverse shock radius can be estimated using Eq.~(\ref{brss}).

The bubble/ISM direct interaction has a strong resemblance 
to the interaction between a supersonic stellar wind and the ISM.
Therefore, for this case 
the solution for a supersonic wind with continuous injection has been adopted (Castor et
al. \cite{cas75}). This renders the ISM forward shock location at
\be
R_{\rm b}\approx 0.76\,\left(\frac{L_{\rm sd}}{\rho_{\rm
ism}}\right)^{1/5}\,t_{\rm p}^{3/5}\,,
\label{rbsol}
\ee
moving with velocity
\be
v_{\rm b}\approx (3/5)\,R_{\rm
b}/t_{\rm p}\,.
\ee
The inner pressure is now
\be
P_{\rm b}\sim \rho_{\rm ism}\,v_{\rm b}^2\,,
\ee
and again the preshock and postshock total 
pressure balance gives the location of the bubble reverse shock: $R_{\rm
brs}\approx \sqrt{L_{\rm sd}/2\pi P_{\rm b}v_{\rm exp}}$. Since the bubble 
external medium has a low sound speed, the precise shape of the bubble 
is less clear now, although it has been assumed to be spherical, as discussed 
in Sect.~\ref{sfe}.

For old enough sources, say $t_{\rm p}>10^5$~yr, 
the proper motion of the system can affect the ISM shock, leading to
the formation of a bow-shaped ISM shock. In this case, the reverse shock is located
approximately at the point where the bubble and ISM ram pressures become equal:
\be
R_{\rm brs}\sim \sqrt{\frac{L_{\rm sd}}{2\pi v_{\rm exp}\rho_{\rm ism}v_{\rm p}^2}}\approx 
0.3 \mbox{ pc}\,.
\label{rspm}
\ee

\begin{figure}[!h]
   \centering
\includegraphics[angle=0, width=0.5\textwidth]{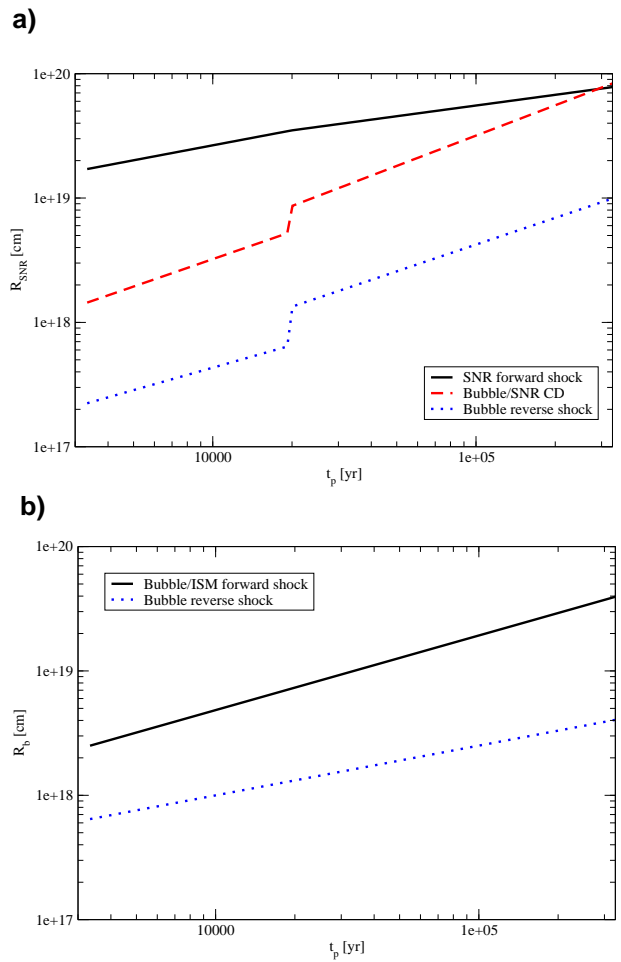}
\caption{{\bf a)} Evolution with time of the radii of the SNR/ISM forward shock (black solid
line), 
SNR/bubble contact discontinuity (red dashed line), and bubble reverse shock
(blue dotted line). The CD radius becomes larger than the SNR shock radius, which
indicates that the model adopted here does not apply after a few $\times 10^5$~yr.
{\bf b)} Evolution with time of the radii of the bubble/ISM forward (black solid
line) and reverse shocks (blue dotted line).}
\label{f1}
\end{figure}

\begin{figure}[!h]
   \centering
\includegraphics[angle=0, width=0.5\textwidth]{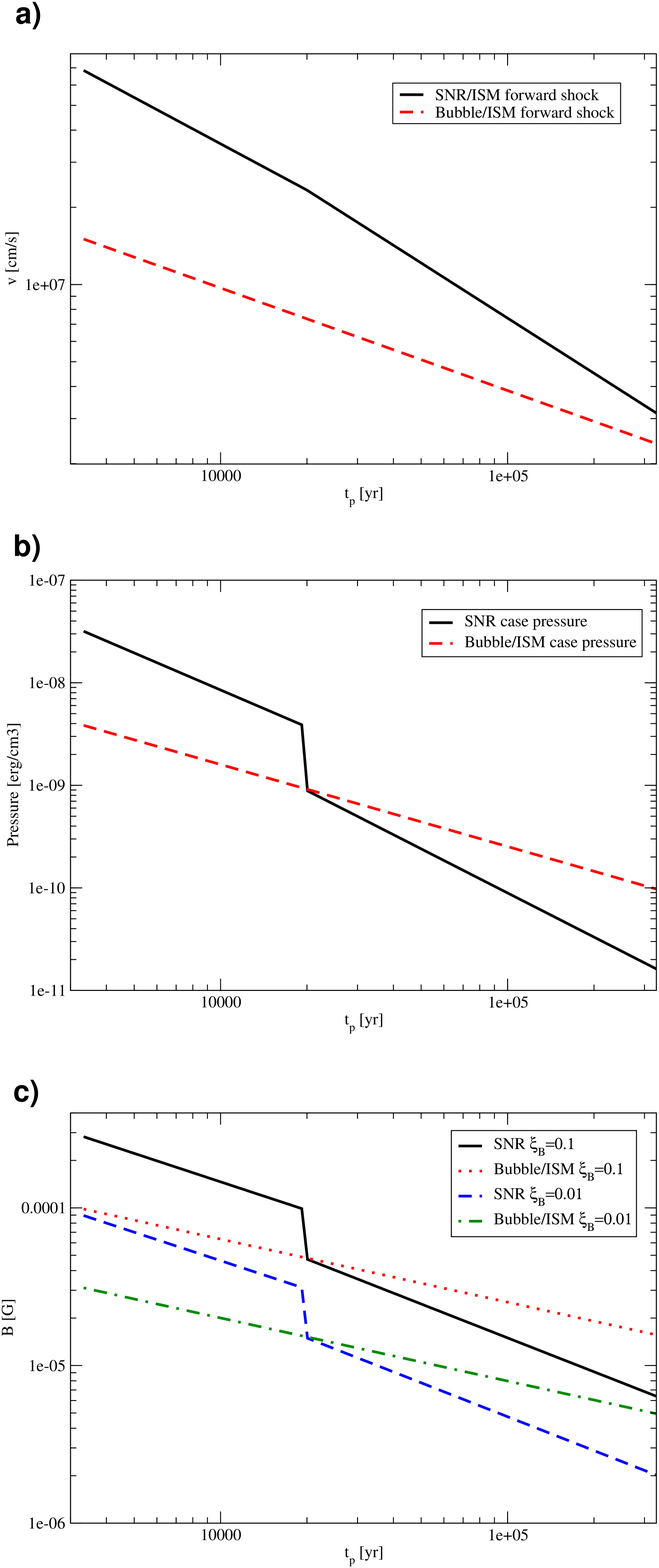}
\caption{{\bf a)} Evolution with time of the SNR/ (black solid line) 
and bubble/ISM forward shock velocities (red dashed line). 
{\bf b)} Evolution with time of
the 
SNR/ (black solid line) and bubble/ISM pressures (red dashed line). {\bf c)}
Evolution with time of the 
SNR/ (black solid/blue dashed line) and bubble/ISM magnetic fields (red dotted/green dot-dashed line).}
\label{f2}
\end{figure}

Using the model just described, the relevant distances, velocities, and pressures can be computed for different system ages.
Parameter values similar to those of LS~5039 and  LS~I~+61~303 have been adopted, and their values are presented in Table~1.
The dynamical, as well as the radiative (see next section), results may also apply to PSR~B1259$-$63, although the comparison
is less straightforward, as discussed in Sect.~\ref{disc}. It is noteworthy that, as long as $x_0$ is much smaller than
the largest scales of interaction with the medium, the size difference between PSR~B1259$-$63 and the other two sources should not 
introduce strong differences in the bubble propagation and termination. The results are shown in Figs.~\ref{f1} and \ref{f2}. The breaks in
reverse shock and CD distances, and pressures, apparent in the figures for the SNR case at $2\times 10^4$~yr, are caused by the significant loss
of internal energy (a fraction 1-$\epsilon$) that takes place in the adiabatic-to-radiative phase transition (Blinnikov et al. \cite{biu82}).

The approach carried out here should be reasonable at a semi-quantitative level, but there are some caveats. First, these different
stages of the bubble evolution are idealizations, and mixed situations could take place.
Also, the
initial conditions of the bubble were oversimplified, assuming that most of the complexity of the interaction on the
binary scales is smoothed out. In particular, the importance of the geometry of the interacting outflows (e.g. polar
winds, disk, etc.) is to be studied in detail. On larger scales, the shocked bubble, SNR and ISM media were
approximated as homogeneous static gas. Rayleigh-Taylor instabilities in the CD between the denser shocked ISM shell, and
its inner region, were neglected, along with anisotropies in the ISM. To account for all this properly,
magnetohydrodynamical simulations of the bubble formation, evolution and termination are planned. We
also neglected the pulsar spin-down luminosity evolution with time, which may imply an underestimate of the total bubble
injected energy by a factor of a few. 

\begin{table}{
\label{tab}
\begin{center} 
  \begin{tabular}{lcc}
    \hline
Parameter [units] & value \\
     \hline\\
Star mass-loss rate [$M_\odot$~yr$^{-1}$] & $3\times 10^{-7}$ \\
Stellar wind velocity [cm~s$^{-1}$] & $2\times 10^8$ \\
Pulsar spin-down luminosity [erg~s$^{-1}$] & $10^{37}$ \\
SNR energy [erg] & $10^{51}$ \\
ISM number density [cm$^{-3}$] & 10 \\
Magnetic-to-total energy density ratio & 0.01, 0.1 \\
Nonthermal-to-total luminosity ratio & 0.01 \\ 
Star luminosity [erg~s$^{-1}$] & $10^{39}$ \\
Stellar photon energy [eV] & 10 \\
Reverse shock velocity [cm~s$^{-1}$] & $10^9$ \\
Source distance [kpc] & 3 \\
  \end{tabular}
\end{center}
\caption{Adopted parameter values}
}
\end{table}

\section{Nonthermal radiation estimates}\label{nt}

The reverse shock is a good candidate for nonthermal emission, since it is fast and strong, and basically all the bubble
energy goes through it. The forward shock in the ambient medium, on the other hand, is either much slower and also less energetic
(bubble/ISM case), or just a sound wave (SNR case). 

\subsection{Emitter characterization}

To compute the particle acceleration rate in the reverse shock, we adopted the expression for a non-relativistic
hydrodynamical shock in the test particle and Bohm diffusion approximations: $\dot{E}=(1/2\pi)(v_{\rm exp}/c)^2\,q\,B\,c$
(e.g. Drury \cite{dru83}), where $B$ is the magnetic field and $q$ the particle charge. The same $B$ and diffusion
coefficient values were assumed in both sides of the shock. The resulting energy distribution of the particles injected
in the emitter depends on energy as $\propto E^{-2}$. The $B$-energy density was taken equal to $\xi_{\rm B}\,P_{\rm
SNR|b}$, with $\xi_{\rm B}=0.1$ and 0.01. Smaller values seem less plausible, because some magnetic field is expected to be
transported from the binary scales, but cannot be discarded. The maximum energies are the minimum value between the
synchrotron-limited case, $E_{\rm max}\approx \sqrt{q\,c/a_{\rm s}\,B}$, with $a_{\rm s}=1.6\times 10^{-3}$, and the
diffusive-escape one, $E_{\rm max}\approx \sqrt{3/4\pi}\,(v_{\rm b}/c)\,q\,B\,R_{\rm brs}$. Although the stellar
photon energy density, the dominant radiation field, is higher than the $B$-energy density at $R_{\rm brs}$, the Klein-Nishina (KN) effect makes IC losses
less important than synchrotron (or escape) ones at $E_{\rm max}$. 
For a wide source age range, in the case of electrons $E_{\rm max}$ is $\sim
100$~TeV, but determined by synchrotron cooling for $\xi_{\rm B}=0.1$, and by
diffusive escape for $\xi_{\rm B}=0.01$.  For protons, the maximum energies are $\sim 300$~TeV for $\xi_{\rm B}=0.1$ and $100$~TeV for
$\xi_{\rm B}=0.01$, both limited by diffusive escape. The differences between the maximum energies in the SNR and the
bubble/ISM case are small; in particular, under diffusive escape, they are equal. Synchrotron-limited maximum energies
are $\propto B^{-1/2}$, but diffusive-escape ones are constant, because $B\times R_{\rm brs}$ is also constant.

The conditions downstream of the bubble reverse shock, in particular the low densities, render hadronic processes
inefficient. Therefore, we have focused on electrons emitting via synchrotron and inverse Compton (IC) scattering. The most
energetic protons may diffuse away from the reverse shock to reach the shocked ISM, but to be efficient, this mechanism
requires target densities to be much higher than those considered here. To compute the synchrotron and IC emission (e.g. Blumenthal
\& Gould \cite{blu70}), we modeled the emitter as only one zone, taking homogeneous $B$- and radiation fields. This
approach is valid for the synchrotron calculations as long as the flow is subsonic, and $\xi_{\rm B}$ is the same everywhere,
which is admittedly a rough approximation in an ideal MHD flow. The role of diffusion in electron transport is negligible in
the present context unless diffusion coefficients are much higher than Bohm. Owing to fast cooling, synchrotron X-rays are
radiated close to the reverse shock region, whereas the synchrotron radio emission comes from a more distant location due to
advection. The particle flux conservation means that the advection speed downstream of the reverse shock is roughly $\propto 1/R^2$,
where $R$ is the distance to the reverse shock. This makes particles accumulate at
\be R\sim R_{\rm brs}\,(1+3\,v_{\rm
exp}\,t_{\rm cool}/4R_{\rm brs})^{1/3}\sim 10\,R_{\rm brs}\,,  
\ee 
with $t_{\rm cool}$ the relevant cooling time. Most
of the IC radiation comes also from a distance $\sim R$ to the star. This is not
true for very high-energy IC photons, but for the adopted $B$-values and due to the KN effect, their fluxes are minor so were 
neglected here (see however Sect.~\ref{disc}). Given the quasi-spherical shape of the emitter, the target photon
field for IC has been taken as isotropic and monoenergetic, given its narrow band. The stellar luminosity and photon energy
have been fixed to $10^{39}$~erg~s$^{-1}$ and 10~eV. The luminosity injected in the form of
nonthermal electrons at the bubble reverse shock was taken as 1\% of the pulsar wind luminosity: 
$L_{\rm nt}=0.01\,L_{\rm sd}$. Although $L_{\rm nt}$ is hard to determine, changes in this parameter only affect the nonthermal fluxes linearly.
Adiabatic cooling has been included in the calculations, with the cooling timescale taken as $R_{\rm SNR|b}/v_{\rm SNR|b}$.

\subsection{Spectral energy distributions and lightcurves}

After characterizing the emitter as a region with roughly homogeneous properties, the maximum particle energies, and the
dominant cooling processes (adiabatic and radiative), it is easy to calculate the electron evolution. For that, a
constant particle injection is assumed, and electrons are evolved up to the source age of interest. Once the particle
energy distribution is known, the synchrotron and IC spectral energy distributions, as well as the specific and
integrated fluxes, are calculated by convolving the synchrotron and IC specific power per electron by the particle energy distribution.

\begin{figure}[!h]
   \centering
\includegraphics[angle=270, width=0.55\textwidth]{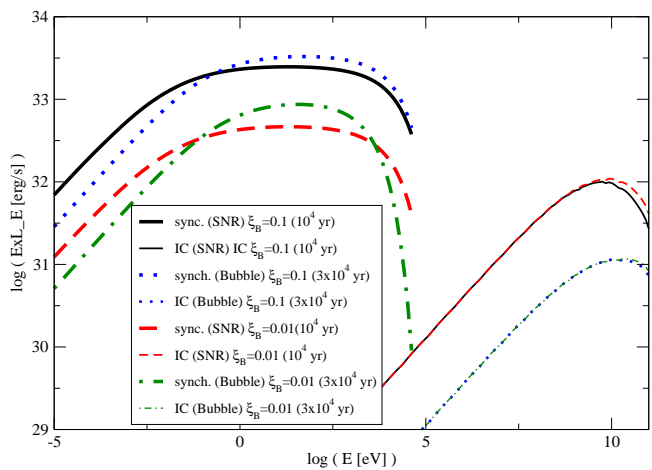}
\caption{Spectral energy distributions of the synchrotron (thick lines) and IC
(thin lines) 
emission computed for the SNR ($10^4$~yr) and the bubble/ISM ($3\times 10^4$~yr) cases, and $\xi_{\rm
B}=0.01,\,0.1$.}
\label{f4}
\end{figure}

\begin{figure}[!h]
   \centering
\includegraphics[angle=0, width=0.5\textwidth]{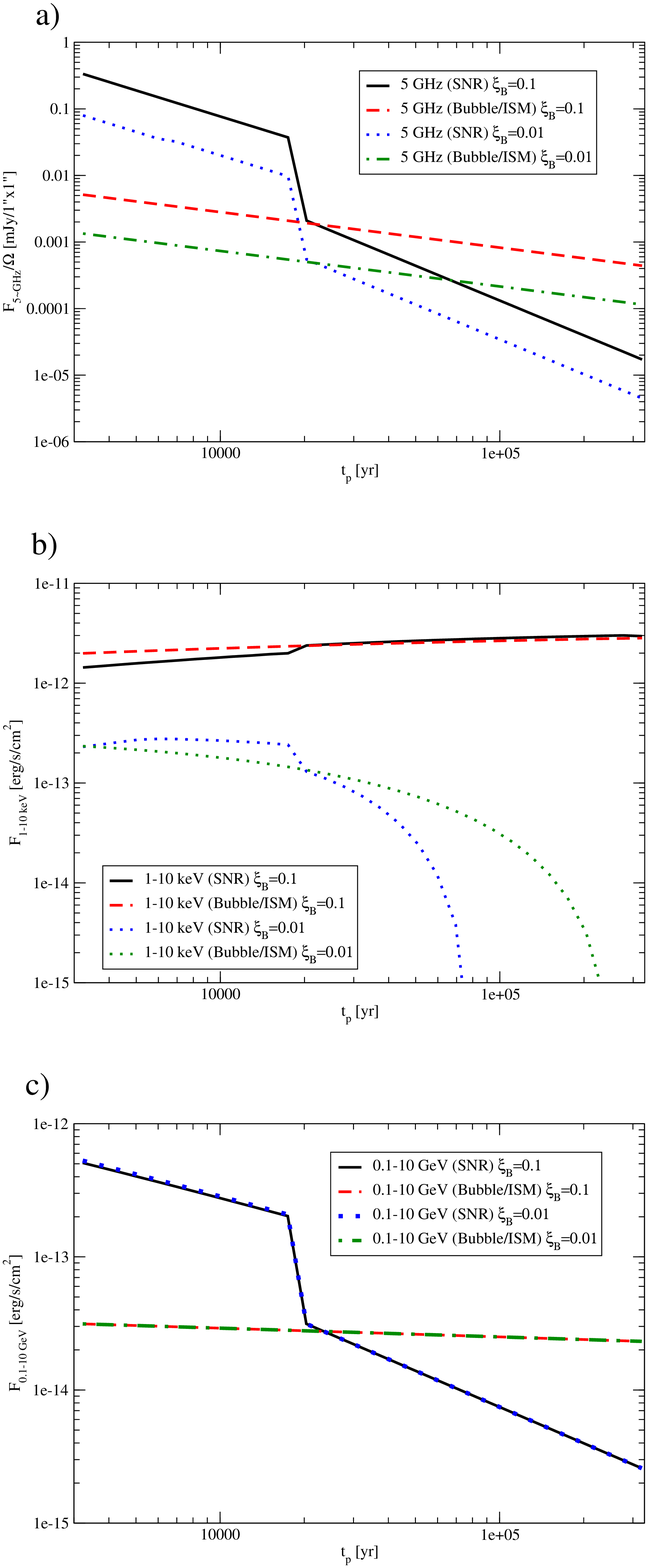}
\caption{{\bf a)} Lightcurve of the radio (5~GHz) surface brightness. {\bf b)} Lightcurve 
of X-ray flux in the range 1--10~keV. {\bf c)} Lightcurve of the IC flux in the range 0.1--10~GeV. 
Both the SNR and the bubble/ISM cases are shown in the age range $t_{\rm p}=3\times
(10^3-10^{5})$~yr and for $\xi_{\rm B}=0.01,\,0.1$.}
\label{f5}
\end{figure}

The results of the radiation calculations are shown in Figs.~\ref{f4} and \ref{f5}. Figure~\ref{f4} shows the computed synchrotron and IC spectral energy distributions for a source age of
$10^4$~yr (SNR) and $3\times 10^4$~yr (bubble/ISM), with the remaining parameter values given in Table~1. Figure~\ref{f5} presents the time evolution of the surface brightness at 5~GHz, and
the 1-10~keV and 0.1-10~GeV bolometric fluxes. The radio surface brightness was computed by just dividing the specific flux by the source solid angle in  $1"\times 1"$ units at 3~kpc,
taking $R$ as the source size.  We point out that our aim is just to provide with a crude estimate of the expected radiation. More detailed calculations of the nonthermal emission will be
presented elsewhere. 

As seen in the figures, the break of the adiabatic-to-radiative phase transition appears in all the SNR lightcurves, although is more apparent for the radio surface brightness and IC
fluxes, which are strongly affected by the jump in $R$. The general evolution of $R$ produces the long-term strong decay of the SNR radio and GeV lightcurves, and a smoother decay of the
bubble/ISM radio lightcurve, whereas the bubble/ISM GeV lightcurve remains more or less constant. The differences are partially caused by the evolution of the relative importance of the
different radiation channels and the adiabatic one, which is not the same in the SNR and the bubble/ISM cases. The synchrotron and IC emissivities have also different time dependencies, the
former depending on $\xi_{\rm B}$ as well. For $\xi_{\rm B}=0.1$, X-ray emitting electrons radiate all their energy through synchrotron, rendering a flat X-ray lightcurve and photon indexes
$\sim 2$. For $\xi_{\rm B}=0.01$, with the electron maximum energies limited by diffusive-escape, the maximum energy of synchrotron photons ($\propto B\,E^2$) decreases with time, and it
quenches the X-ray flux for $t_{\rm p}\ga 10^5$~yr. This also explains the steepening of the synchrotron X-ray spectrum for the bubble/ISM case seen in Fig.~\ref{f4}. The SNR X-ray spectrum
does not show this feature because it has been computed for $t=10^4\,{\rm yr}<t_{\rm c}$.

\section{Discussion}\label{disc}

The sizes of the bubble reverse shock and its downstream region give an idea of the source angular size at different wavelengths. At 3~kpc, the hard X-ray emitter would have $\theta\approx
10"$ when still inside the SNR ($t\la 10^4$~yr), and $\sim 20"-1'$ in the bubble/ISM case ($t\approx 10^4-10^5$~yr). These $\theta$ values were calculated for the $\rho_{\rm ism}$, $E_{\rm
SNR}$, and $L_{\rm sd}$ values given in Table~1, although we need to recall their weak inter-dependence: $\theta\propto (E_{\rm SNR}\,|\,L_{\rm sd}/\rho_{\rm ism})^{1/5}$.  Extended X-ray
fluxes could be detectable by an instrument like {\it Chandra} up to $t_{\rm p}\sim 10^5$~yr. It is worth noting that, if synchrotron X-ray energies are detected from the reverse shock,
the diffusion coefficient should be close to Bohm. 

Radio emission is possibly detectable for the SNR case with $\xi_{\rm B}=0.1$, it may be hard to be found for the case $\xi_{\rm B}=0.01$, and its detection seems discarded for the
bubble/ISM case for any $\xi_{\rm B}$-value.

The GeV fluxes seem too low to be detectable by any current instrument, and the chances are even lower in TeV due to the
Klein-Nishina effect and dominant synchrotron cooling. However, it may still be possible to get higher IC GeV and TeV fluxes
increasing ten times $L_{\rm nt}$, and reducing strongly $\xi_{\rm B}$ not to overcome the radio/X-ray constraints.
This would lead to GeV-TeV fluxes that could be detectable by present (GeV: {\it Fermi}, {\it AGILE}; TeV: HESS, MAGIC, VERITAS) and/or
forthcoming instruments (e.g. TeV: CTA). Given the spectral break at $\sim 10$~GeV introduced by adiabatic losses, the bubble
reverse shock would be a good target for a $\sim 10$~GeV-threshold Cherenkov instrument. 

We now briefly discuss whether the evidence or hints of extended X-rays from LS~5039, LS~I~+61~303, and PSR~B1259$-$63
can be understood in the scenario posed here. 

\subsection{LS~5039}

LS~5039 is likely a young source with an age range $(4-10)\times 10^4$~yr, and it has already abandoned its SNR (e.g. Rib\'o et al.
\cite{rib02}; Mold\'on et al. \cite{mol11c}). The angular size at 2.5~kpc for the bubble/ISM case is consistent with the
observed angular size of the emitter, $\theta\sim 1'-2'$, and the fluxes can be explained in the framework given here. The
photon index at distances $\sim 1'$ is $\sim 1.9\pm 0.7$, consistent with the SED shown in Fig.~\ref{f4}, and then softens
farther out reaching $\sim 3.1\pm 0.5$ at $\sim 2'$ (Durant et al. \cite{dur11}). This can be explained by the cooling of
the highest energy electrons close to the reverse shock, thereby depleting the highest energy part of photons and softening the
spectrum farther from the shock. The adopted $n_{\rm ism}=10$~cm$^{-3}$ is similar to the values derived for the
surroundings of the LS~5039 SNR candidate (Rib\'o et al. \cite{rib02}), and the pulsar $L_{\rm sd}$ is probably $\sim
10^{37}$~erg~s$^{-1}$, which is required to explain the GeV luminosity of the source (see Sect.~4 in Zabalza et al \cite{zab11a}). The
morphology of the extended radio emission found  by Durant et al. (\cite{dur11}) is not completely spherical, which may be
explained by the only partial isotropization  of the momentum flux in the inner regions of the bubble (see Sect.~\ref{sfe}). 
A scenario with $\eta>1$ could be also behind the anisotropy (see Sect.~\ref{pulsd}), but it seems less likely.

\subsection{LS~I~+61~303}

Assuming that the hints of extended X-rays in LS~I~+61~303 are real, we see that the angular size, $\theta\sim 10"$, is
compatible with the bubble/SNR scenario for $t_{\rm p}\la 10^4$~yr, but with a low nonthermal efficiency or a very low
magnetic field, given the weak X-ray flux (Paredes et al. \cite{par07}) and the radio nondetection on those scales (Mart\'i et al. \cite{mar98}). 

The SNR progenitor of LS~I~+61~303 has not yet been found. However, extended 5~GHz emission has been detected centered on the
location of  LS~I~+61~303, with a typical size of 6--8 arcminutes and fluxes of tens of mJy (Mart\'i et al. \cite{mar98};
Mu\~noz-Arjonilla et al. \cite{mun09}). It has been argued that the lack of extended X-rays on the same scales, and the low
surface radio brightness (in a nonthermal scenario), may go against an SNR interpretation (Mart\'i et al. \cite{mar98};
Mu\~noz-Arjonilla et al. \cite{mun09}). However, the radio emission surrounding LS~I~+61~303 may be mostly thermal
(Mu\~noz-Arjonilla et al. \cite{mun09}), coming from an SNR/ISM shock close to its radiative phase. If the $n_{\rm ism}$
values in the region of LS~I~+61~303 were a bit higher than those adopted here, an SNR remnant with energy $E_{\rm SNR}\sim
10^{51}$~erg would enter its radiative phase after $\sim 10^4$~yr, with a radius $\sim 10^{19}$~cm, a few arcminutes at
2~kpc. Thermal X-rays would not be expected in that case, since the shocked ISM
material would be too cold. The slow proper motion of the source (Dhawan et al. \cite{dha06}) is compatible with LS~I~+61~303
being at the center of the SNR (see however Mart\'i et al. \cite{mar98} for image deformation). As in LS~5039, the $L_{\rm
sd}$-value in LS~I~+61~303 is expected to be $\sim 10^{37}$~erg~s$^{-1}$ because of the high GeV luminosity of the
source (see Sect.~5 in Zabalza et al \cite{zab11b}).

\subsection{PSR~B1259$-$63}

PSR~B1259$-$63 has an age of $t_{\rm p}\approx 3\times 10^5$~yr (Wex et al. \cite{wex98}), and is likely now interacting directly with
the ISM. With this age, and the pulsar spin-down luminosity $L_{\rm sd}\approx 8\times 10^{35}$~erg~s$^{-1}$ (Manchester
et al. \cite{man95}), Eq.~(\ref{rbsol}) yields $R_{\rm brs}\sim 2\times 10^{18}$~cm ($\sim$~1'), assuming $n_{\rm ism}=10$~cm$^{-3}$.
This is about one arcminute at 2.3~kpc and much larger than the observed $\theta\sim 4"-15"$ (Pavlov et al. \cite{pav11a}).
Nevertheless, in PSR~B1259$-$63 it seems likely that the proper motion has already bow-shaped the ISM shock. In that case,
for $n_{\rm ism}=10$~cm$^{-3}$ and $v_{\rm p}\sim 10^7$~cm~s$^{-1}$, the reverse shock would be located at a distance of
$\sim 3\times 10^{17}$~cm from the system, or $\sim 10"$ neglecting projection effects. Unfortunately, the large errors
of the proper motion measurements (Zacharias et al. \cite{zac09}) and the low statistics of the X-ray data make it difficult
to compare the X-ray extension and proper motion directions. 

There is another possibility behind the extended X-rays found in PSR~B1259$-$63. The most significant detection comes from
$\sim 4"$, or $\sim 10^{17}$~cm in linear (projected) size, which is $\sim v_{\rm exp}\,T$. Although the flow bending
distance, $x_0$, should be less than that, the X-ray emission may be related to the asymmetric shock produced by the
Coriolis force, and/or to spiral-arm merging. This radiation could not be resolved in the case of LS~I~+61~303 and LS~5039, which are much
more compact than PSR~B1259$-$63.

\section{Final remarks}\label{rem}

Gamma-ray binaries hosting powerful pulsars can produce supersonic flows that originate in the interaction of a pulsar
wind with the wind of the companion, which may present strong anisotropies and inhomogeneities. The likely mixing of these
flows on larger scales than the binary will render an expanding, rather isotropic, supersonic bubble, which will
eventually terminate in the surrounding medium. The interaction of the supersonic bubble with the surrounding medium can take
place in different ways depending on the age of the pulsar. For young sources, the environment will be the hot SNR ejecta,
whereas it will be the shocked ISM for older objects. Although this interaction is expected to be rather symmetric, for old
enough sources, possibly like PSR~B1259$-$63, the proper motion can bow-shape the interaction structure with the ISM. 

The radiation from the shocked bubble interacting with the environment may explain the observed extended X-ray emission from LS~5039, and possibly also from LS~I~+61~303 (if real). For PSR~B1259$-$63, the found extended X-rays could come from inner regions of the bubble, triggered perhaps by Coriolis force shocks or some other type of dissipation mechanism.
Extended X-ray emission  may also eventually be detected in HESS~J0632$+$057 and 1FGL~J1018.6$-$5856, which could also originate in shocked wind outflows. The shape of the interaction
region might allow the distinction of the driving flow, i.e. a jet or a pulsar wind, but as shown in Sect.~\ref{stardom} it may be not straightforward.

The complexity of the flow evolution on the different relevant scales makes a proper analytical characterization difficult, and thus makes 
magnetohydrodynamical simulations important.
The geometry, level of anisotropy, velocity and density of the stellar flow requires careful study, in particular for binaries hosting 
stars with an equatorial disk, like PSR~B1259$-$63 and
LS~I~+61~303. Anisotropy in the pulsar wind has been also neglected here, but may play some role, at least up to intermediate scales.

\begin{acknowledgements}
The authors want to thank Dmitry Khangulyan for thoroughly reading the manuscript, and for his useful 
comments and suggestions. The authors are grateful to the referee, Ignacio Negueruela, for many constructive
and useful comments.
The research leading to these results has received funding from the European
Union
Seventh Framework Program (FP7/2007-2013) under grant agreement
PIEF-GA-2009-252463.
V.B.-R. acknowledges support by the Spanish Ministerio de Ciencia e 
Innovaci\'on (MICINN) under grants AYA2010-21782-C03-01 and 
FPA2010-22056-C06-02.
B.M.V. thanks to State contract 2011-1.4-508-008/9 from FTP of RF Ministry 
of Education and Science.
\end{acknowledgements}


\begin{thebibliography}{}
\small{
\bibitem[2009a]{abd09a} Abdo, A.~A., et al. 2009a, 2009, Science, 326, 1512
\bibitem[2009b]{abd09b} Abdo, A.~A., et al. 2009b, ApJ, 701, L123
\bibitem[2009c]{abd09c} Abdo, A.~A., et al. 2009c, ApJ, 706, L56
\bibitem[2010a]{abd10a} Abdo, A.~A., et al. 2010a, Science, 329, 817
\bibitem[2010b]{abd10b} Abdo, A. A., et al. 2010b, \apj, 723, 649
\bibitem[2011]{abd11} Abdo, A. A., et al. 2011, \apj, 736, L11
\bibitem[2005a]{aha05a} Aharonian, F.~A. et al. 2005a, A\&A, 442, 1
\bibitem[2005b]{aha05b} Aharonian, F.~A. et al. 2005b, Science, 309, 746
\bibitem[2007]{aha07} Aharonian, F.~A. et al. 2007, A\&A, 469, L1
\bibitem[2006]{alb06} Albert, J. et al. 2006, Science, 312, 1771 
\bibitem[2007]{alb07} Albert, J. et al. 2007, \apj, 665, L51
\bibitem[1982]{biu82} Blinnikov, S. I., Imshennik, V. S., Utrobin, V. P. 1982, SvAL, 8, 361
\bibitem[1970]{blu70} Blumenthal, G. R., Gould, R. J., 1970, Rev. Mod. Phys., 42, 237
\bibitem[2002]{bog02} Bogovalov, S. V., \& Khangoulian, D. V. 2002, MNRAS, 336, L53
\bibitem[2008]{bog08} Bogovalov, S.~V., Khangulyan, D., Koldoba, A.~V.,
Ustyugova, G.~V., Aharonian, F.~A. 2008, MNRAS, 387, 63
\bibitem[2011]{bog11} Bogovalov, S., Khangulyan, D., Koldoba, A. V., Ustyugova, G. V., Aharonian, F.A. 2011, MNRAS, submited [astro-ph/1107.4831]
\bibitem[1944]{bon44}
Bondi, H. \& Hoyle, F.,1944, MNRAS, 104, 273
\bibitem[2011]{bon11}	
Bongiorno, S. D., Falcone, A. D., Stroh, M., et al. 2011, ApJ, 737, L11
\bibitem[2009]{bor09} Bordas, P., Bosch-Ramon, V., Paredes, J. M., Perucho, M.
2009, A\&A, 497, 325
\bibitem[2009]{bos09} Bosch-Ramon, V. \& Khangulyan, D. 2009, IJMPD, 18, 347
\bibitem[2005]{bos05} Bosch-Ramon, V., Aharonian, A.~F., Paredes, J.~M. 2005,
A\&A, 432, 609
\bibitem[2006]{bos06} Bosch-Ramon, V., Paredes, J.~M., Romero, G. E., Rib\'o, M.
2006, A\&A, 459, L25
\bibitem[2008]{bos08} Bosch-Ramon, V., Khangulyan, D., Aharonian, F. A. 2008,
A\&A, 489, L21
\bibitem[2011]{bos11} Bosch-Ramon, V., Perucho, M., Bordas, P., 2011, A\&A, 528,
89		
\bibitem[1975]{cas75} Castor, J., McCray, R., Weaver, R. 1975, \apj, 200, L107
\bibitem[2005a]{cas05a} Casares, J., Rib\'o, M., Ribas, I., et al., 2005a,
MNRAS, 364, 899
\bibitem[2005b]{cas05b} Casares, J., Ribas, I., Paredes, J. M., Mart\'i, J.,
Allende Prieto, C. 2005b, MNRAS, 360, 1105
\bibitem[2006]{che06} Chernyakova, M., Neronov, A., Walter, R. 2006, MNRAS, 372,
1585
\bibitem[2001]{cla01} Clark, J.~S., Reig, P., Goodwin, S.~P. 2001, A\&A, 376,
476
\bibitem[2011]{cor11} Corbet, R.~H.~D., Cheung, C.~C., Kerr, M., 2011, ATel,
3221, 1
\bibitem[1972]{cox72} Cox, D.P., 1972, \apj, 178, 159
\bibitem[2010]{deo10} de Ona Wilhelmi, E. 2010, 38th COSPAR Scientific Assembly,
Bremen, Germany, p.6
\bibitem[2006]{dha06} Dhawan, V., Mioduszewski, A., Rupen, M. 2006, Proceedings
of the VI Microquasar Workshop: Microquasars and Beyond, Como, Italy, 52.1
\bibitem[1983]{dru83} Drury, L.O'C., 1983, Reports on Progress in Physics, 46,
973
\bibitem[2006]{dub06} Dubus, G. 2006, A\&A, 456, 801 
\bibitem[2011]{dur11} Durant, M., Kargaltsev, O., Pavlov, G.~G., et al. 2011,
\apj, 735, 58
\bibitem[2010]{fal10} Falcone, A. D., Grube, J., Hinton, J., et al. 2010, \apj,
708, L52
\bibitem[1991]{fra91}Frail, D. A., \& Hjellming, R. M. 1991, AJ, 101, 2126
\bibitem[2005]{gal05} Gallo, E., Fender, R., Kaiser, C., et al. 2005, Nature,
436, 819	
\bibitem[2003]{hei03} Heindl, W. A., Tomsick, J. A., Wijnands, R., Smith, D. M.
2003, \apj, 588, L97
\bibitem[2002]{hei102} Heinz, S., 2002, A\&A, 388, L40
\bibitem[2002]{hei202} Heinz, S. \& Sunyaev, R. 2002, A\&A, 390, 751
\bibitem[2011]{hil11} Hill, A. B., Szostek, A., Corbel, S., et al. 2011, MNRAS, 415, 235 
\bibitem[2009]{hin09} Hinton, J. A., Skilton, J. L., Funk, S., et al. 2009, ApJ,
690, L101	
\bibitem[1981]{hut81} Hutchings, J. B., Crampton, D. 1981, PASP, 93, 486
\bibitem[2008]{kha08} Khangulyan, D. V., Aharonian, F. A., Bogovalov, S. V.,
Koldoba, A. V., Ustyugova, G. V. 2008, IJMPD, 17, 1909
\bibitem[1992]{joh92} Johnston, S., Manchester, R.~N., Lyne, A. G., et al. 1992,
\apj, 387, L37
\bibitem[2011]{mcs11} McSwain, M. V, Ray, P. S., Ransom, S. M., et al. 2011, \apj, 738, 105
\bibitem[1995]{man95} Manchester, R. N., Johnston, S., Lyne, A. G., D'Amico, N.,
Bailes, M., Nicastro, L. 1995, ApJ, 445, L1374
\bibitem[1981]{mar81} Maraschi, L. \& Treves, A. 1981, MNRAS, 194, 1
\bibitem[1996]{mar96} Mart\'i, J., Rodr\'iguez, L.~F., Mirabel, I.~F., Paredes,
J. M. 1996, A\&A, 306, 449
\bibitem[1998]{mar98} Mart\'i, J., Peracaula, M., Paredes, J. M., Massi, M.,
Estalella, R. 1998, A\&A, 329, 951
\bibitem[2005]{mar05} Martocchia, A., Motch, C., Negueruela, I., 2005, A\&A,
430, 245 
\bibitem[2001]{mas01} Massi, M., Rib\'o, M., Paredes, J. M., Peracaula, M.,
Estalella, R. 2001, A\&A, 376, 217 
\bibitem[2004]{mas04} Massi, M., Rib\'o, M., Paredes, J. M., et al. 2004, A\&A,
414, L1 
\bibitem[2009]{mas09} Massi, M., Kaufman Bernad\'o, M. 2009, \apj, 702, 1179
\bibitem[2011a]{mol11a} Mold\'on, J., Johnston, S., Rib\'o, M., Paredes, J.~M.,
Deller, A. T. 2011a, ApJ, 732, L10
\bibitem[2011b]{mol11b} Mold\'on, J., Rib\'o, M., Paredes, J.~M. 2011b, A\&A, 533, L7
\bibitem[2011c]{mol11c} Mold\'on, J., Ribo, M., Paredes, J. M. 2011c,
High-Energy Emission from Pulsars and their Systems, Astrophysics and Space Science Proceedings, 
ISBN 978-3-642-17250-2. Springer-Verlag Berlin Heidelberg, 2011, p. 1-2
\bibitem[2009]{mun09} Mu\~noz-Arjonilla, A. J., Mart\'i, J., Combi, J. A. 2009,
A\&A, 497, 457
\bibitem[2011]{neg11} Negueruela, I., Rib\'o, M., Herrero, A., Lorenzo, J.,
Khangulyan, D., Aharonian, F.~A. 2011, \apj, 732, L11
\bibitem[2011]{oka11} 
Okazaki, A. T., Nagataki, S., Naito, T., et al. 2011, PASJ, in press [astro-ph/1105.1481]
\bibitem[2000]{par00} Paredes, J. M., Mart\'i, J., Rib\'o, M., \& Massi, M.
2000, Science, 288, 2340	
\bibitem[2002]{par02} Paredes, J. M., Rib\'o, M., Mart\'i, J., 2002, A\&A, 393,
99 
\bibitem[2006]{par06} Paredes, J.~M., Bosch-Ramon, V., Romero, G.~E., 2006,
A\&A, 451, 259 
\bibitem[2007]{par07} Paredes, J. M., Rib\'o, M., Bosch-Ramon, V., et al. 2007,
ApJ, 664, L39
\bibitem[2003]{por03} Porter, J. M. \& Rivinius, T. 2003, PASP, 115, 1153
\bibitem[2011a]{pav11a} Pavlov, G.~G., Chang, C., Kargaltsev, O. 2011a, ApJ, 730, 2
\bibitem[2011b]{pav11b} Pavlov, G. G., Misanovic, Z., Kargaltsev, O, Garmire, G.
P. 2011b, ATel, 3228
\bibitem[2009]{pita09} Pittard, J. M. 2009, MNRAS, 396, 1743
\bibitem[2009]{pul09} Puls, J., Sundqvist, J. O., Najarro, F., Hanson, M. M. 2009, AIPC, 1171, 1234 
\bibitem[2010]{rea10} Rea, N., Torres, D. F., van der Klis, M., Jonker, P. G.,
M\'endez, M., Sierpowska-Bartosik, A. 2010, MNRAS, 405, 2206
\bibitem[2011]{rea11}
Rea, N., Torres, D. F., Caliandro, G. A., et al. 2011, MNRAS, 416, 1514 
\bibitem[2002]{rib02} Rib\'o, M., Paredes, J. M., Romero, G. E., et al. 2002,
A\&A, 384, 954
\bibitem[2011]{rib11} Rib\'o, M., 2011, invited talk at HEPROIII, held in Barcelona, Spain, in June 2011
\bibitem[2005]{rom05} Romero, G. E., Christiansen, H. R., Orellana, M. 2005,
\apj, 632, 1093
\bibitem[2007]{rom07} Romero, G.~E., Okazaki, A.~T., Orellana, M., Owocki, S.~P.
2007, A\&A, 474, 15
\bibitem[2007]{rus07} Russell, D.~M., Fender, R.~P., Gallo, E., Kaiser, C. R.
2007, MNRAS, 376, 1341	
\bibitem[2010]{sab10} Sabatini, S., et al. 2010, \apj, 712, L10
\bibitem[2008]{san08} S\'anchez-Sutil, J. R., Mart\'i, J., Combi, J. A., et al.
2008, A\&A,  479, 523
\bibitem[2011]{sar11} Sarty, G.~E., Szalai, T., Kiss, L.~L., et al. 2011, MNRAS,
411, 1293	
\bibitem[1959]{sed59} Sedov, L. I. 1959, Similarity and Dimensional Methods in
Mechanics (New York: Academic Press)
\bibitem[2008]{sie08} Sierpowska-Bartosik, A. \& Torres, D.~F. 2008, Astropart.
Phys., 30, 239
\bibitem[2009]{sie09} Sierpowska-Bartosik, A. \& Torres, D.~F. 2009, \apj, 693,
1462
\bibitem[2009]{ski09} Skilton, J. L., Pandey-Pommier, M., Hinton, J. A., 2009,
MNRAS, 399, 317
\bibitem[1992]{ste92} Stevens, I. R., Blondin, J. M., \& Pollock, A. M. T. 1992,
ApJ, 386, 265
\bibitem[2010]{tam10} Tam, P.~H.~T., Hui, C.~Y., Huang, R.~H.~H., et al. 2010,
\apj, 724, L207	
\bibitem[1992]{tay92} Taylor, A.~R., Kenny, H.~T., Spencer, R.~E., Tzioumis, A.
1992, ApJ, 395, 268
\bibitem[1997]{tav97}  Tavani, M., Arons, J., 1997, ApJ, 477, 439
\bibitem[2009a]{tav09a} Tavani, M., et al. 2009a, ApJ, 698, L142
\bibitem[2009b]{tav09b} Tavani, M., et al. 2009b, Nature, 462, 620
\bibitem[2000]{vel00} Vel\'azquez, P.~F. \& Raga, A.~C. 2000, A\&A, 362, 780
\bibitem[1998]{wex98} Wex, N., Johnston, S., Manchester, R. N., Lyne, A. G.,
Stappers, B. W., Bailes, M. 1998, MNRAS, 298, 997
\bibitem[2010]{wil10} Williams, S. J., Gies, D. R., Matson, R. A. et al. 2010, \apj, 723, L93
\bibitem[2011a]{zab11a} Zabalza, V., Bosch-Ramon, V., Paredes, J. M. 2011, ApJ, submitted
\bibitem[2011b]{zab11b} Zabalza, V., Paredes, J. M., Bosch-Ramon, V. 2011, A\&A, 527, 9
\bibitem[2009]{zac09} Zacharias, N., et al. 2009, VizieR Online Data Catalog,
1315, 0
\bibitem[2008]{zav08} Zavala, J., Vel\'azquez, P.~F., Cerqueira, A.~H., Dubner,
G. M. 2008, MNRAS, 387, 839
}	
\end{thebibliography}
\end{document}